\documentclass[conference]{IEEEtran}
\IEEEoverridecommandlockouts

\usepackage{cite}
\usepackage{amsmath,amssymb,amsfonts}
\usepackage{algorithmic}
\usepackage{graphicx}
\usepackage{textcomp}
\usepackage{xcolor}
\usepackage{siunitx}
\usepackage{enumitem}
\usepackage{booktabs}
\usepackage{flushend}

\usepackage{glossaries}
\newacronym{lora}{LoRA}{Low-Rank Adaptation}
\newacronym{dfwf}{DFWF}{Detecting Fake Without Forgetting}
\newacronym{asv}{ASV}{Automatic Speaker Verification}
\newacronym{tts}{TTS}{Text to Speech}
\newacronym{cnn}{CNN}{Convolutional Neural Network}
\newacronym{LwF}{LwF}{Learning Without Forgetting}
\newacronym{PSA}{PSA}{Positive Sample Alignment}
\newacronym{roc}{ROC}{Receiver Operating Characteristic}

\def\BibTeX{{\rm B\kern-.05em{\sc i\kern-.025em b}\kern-.08em
    T\kern-.1667em\lower.7ex\hbox{E}\kern-.125emX}}
\begin{document}

\title{Freeze and Learn: Continual Learning with Selective Freezing for Speech Deepfake Detection
\thanks{This material is based on research sponsored by the Defense Advanced Research Projects Agency (DARPA) and the Air Force Research Laboratory (AFRL) under agreement number FA8750-20-2-1004. The U.S. Government is authorized to reproduce and distribute reprints for Governmental purposes notwithstanding any copyright notation thereon. The views and conclusions contained herein are those of the authors and should not be interpreted as necessarily representing the official policies or endorsements, either expressed or implied, of DARPA and AFRL or the U.S. Government.
This work was supported by the FOSTERER project, funded by the Italian Ministry of Education, University, and Research within the PRIN 2022 program. This work was partially supported by the European Union under the Italian National Recovery and Resilience Plan (NRRP) of NextGenerationEU, partnership on ``Telecommunications of the Future'' (PE00000001 - program ``RESTART'').}
}
\author{\IEEEauthorblockN{
Davide Salvi\IEEEauthorrefmark{1},
Viola Negroni\IEEEauthorrefmark{1},
Luca Bondi\IEEEauthorrefmark{2},
Paolo Bestagini\IEEEauthorrefmark{1},
Stefano Tubaro\IEEEauthorrefmark{1}
}\vspace{0.35em}
\IEEEauthorblockA{\IEEEauthorrefmark{1} Dipartimento di Elettronica, Informazione e Bioingegneria (DEIB), Politecnico di Milano\\
\IEEEauthorrefmark{2} Bosch Research, USA - Bosch Center for Artificial Intelligence
}
}

\maketitle

\begin{abstract}

In speech deepfake detection, one of the critical aspects is developing detectors able to generalize on unseen data and distinguish fake signals across different datasets.
Common approaches to this challenge involve incorporating diverse data into the training process or fine-tuning models on unseen datasets.
However, these solutions can be computationally demanding and may lead to the loss of knowledge acquired from previously learned data. 
Continual learning techniques offer a potential solution to this problem, allowing the models to learn from unseen data without losing what they have already learned.
Still, the optimal way to apply these algorithms for speech deepfake detection remains unclear, and we do not know which is the best way to apply these algoritms to the developed models.
In this paper we address this aspect and investigate whether, when retraining a speech deepfake detector, it is more effective to apply continual learning across the entire model or to update only some of its layers while freezing others.
Our findings, validated across multiple models, indicate that the most effective approach among the analyzed ones is to update only the weights of the initial layers, which are responsible for processing the input features of the detector.

\end{abstract}


\section{Introduction}

In recent years, significant progress has been made in the speech deepfake detection field~\cite{cuccovillo2022open, amerini2024deepfake}.
Several detectors have been presented, employing diverse model architectures, processing techniques, and approaches, achieving remarkable performance, particularly in controlled scenarios~\cite{yang2024robust, xie2023domain, li2023voice, mari2023all}.
At the same time, speech generation techniques have also advanced, leading to the production of highly realistic fake speech signals that are challenging to discriminate from real ones~\cite{bhagtani2024recent, patel2023deepfake}.
This presents a challenge for existing detectors, which also need to maintain accurate detection capabilities on these new types of data.
To do so, detectors must be continuously updated, considering signals generated from the latest speech synthesis systems. 
Ideally, this would involve retraining the detector from scratch on a combination of its original training dataset and new samples from these unseen algorithms.
However, this approach is often impractical due to computational constraints or the unavailability of the original data.
As an alternative, pre-trained models are often fine-tuned on novel datasets, improving their ability to detect unseen deepfakes.
Nevertheless, this method can lead to a problem known as \textit{catastrophic forgetting}, where detectors become proficient at identifying new data but lose accuracy in discriminating previously learned samples.

To address the issue of catastrophic forgetting, various techniques have been proposed.
These include methods like continual learning and \gls{lora}, which aim to improve the performance of the models across multiple tasks~\cite{ma2021continual, zhang2023you, zhang2024remember, wang2022low, mulimani2024class, dong2024advancing}.
For example, the authors of~\cite{ma2021continual} propose \gls{dfwf}, a regularization-based continual learning method that includes a knowledge distillation loss function to improve the capabilities of the model. Similarly, in~\cite{zhang2023you} adaptive direction modification is used to mitigate catastrophic forgetting.
Despite the relevant contributions brought by these novel methods, their optimal application has been little investigated.
For instance, it remains unclear whether it is more beneficial to use a continuous learning strategy that updates all of a model's layers or to selectively retrain only a few of them.

In this paper we explore this aspect and investigate the effectiveness of updating weights of a limited number of model layers, aiming to understand the optimal strategies for maintaining model performance while adapting to new data.
We do so by considering a pre-trained speech deepfake detector and dividing it into two logical sections.
We retrain it on three distinct datasets, using continual learning with three approaches: updating all model weights, updating only one section while freezing the other, and vice versa.
Finally, we compare the performance of these configurations.
Our analysis is based on the simplistic yet effective assumption that the initial layers of a network are primarily responsible for processing the model input and generating a meaningful representation of it, while the later layers analyze this representation to produce the final output.
Retraining both these sections may not always be advantageous, as limiting the number of weight updates could help 
preserve the model's prior knowledge. 
Our findings validate this intuition, showing that updating only the weights of the first layers of the model while keeping the others frozen leads to a better retention of the model memory.

\section{Proposed Method}
\label{sec:method}

In this paper, we consider the problem of continual learning in the context of speech deepfake detection and investigate the most effective approach for applying continual learning strategies when training the models.

The problem we address is formally defined as follows.
Let us consider a discrete-time speech signal $\mathbf{x}$ sampled with a sampling frequency $f_\text{s}$ and belonging to a class $y \in \{0, 1 \}$, where \num{0} means that it is authentic while \num{1} indicates that it has been synthetically generated.
We consider $\mathbf{x}$ as the input of a speech deepfake detector $\mathcal{M}$, which is trained to estimate its class as $\hat{y} = \mathcal{M}(\mathbf{x})$, where $\hat{y} \in [0,1]$ indicates the likelihood that the signal $\mathbf{x}$ is fake.

\begin{figure}
    \centering
    \includegraphics[width=0.8\columnwidth]{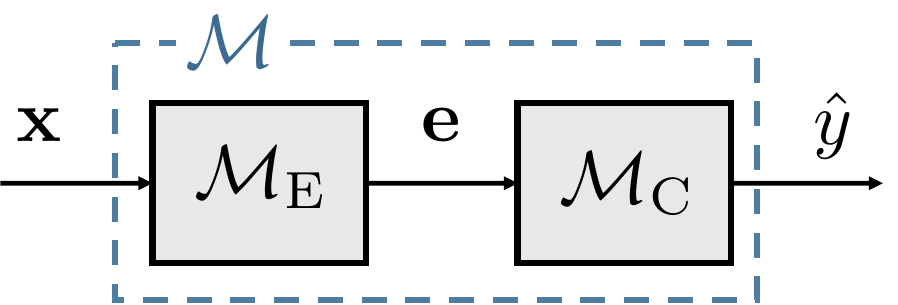}
    \caption{Logical division of a speech deepfake detector $\mathcal{M}$ into its modules.}
    \label{fig:pipeline}
\end{figure}

We consider the detector $\mathcal{M}$ as logically divided into two modules, as it is shown in Figure~\ref{fig:pipeline}:
\begin{itemize}[leftmargin=*]
    \item An \textit{Encoder Module} $\mathcal{M}_\text{E}$ that processes the input $\mathbf{x}$ and maps it into the embedding $\mathbf{e}$, which is a low-dimensional representation of it.
    \item A \textit{Classification Module} $\mathcal{M}_\text{C}$ that transforms the low-dimensional representation $\mathbf{e}$ into the class estimate $\hat{y}$.
\end{itemize}

We train the detector $\mathcal{M}$ on four different datasets $\mathbb{D}_i$, following four distinct training strategies and evaluating how these influence the detection capabilities of the model.
The training strategies we consider are the following:
\begin{itemize}[leftmargin=*]
    \item \textbf{Train-on-All}: The detector $\mathcal{M}$ is simultaneously trained on all the datasets $\mathbb{D}_0$, $\mathbb{D}_1$, $\mathbb{D}_2$, $\mathbb{D}_3$.
    \item \textbf{Fine-tuning}: The detector $\mathcal{M}$ is initially trained on the first dataset $\mathbb{D}_0$ and then fine-tuned sequentially on $\mathbb{D}_1$, $\mathbb{D}_2$, $\mathbb{D}_3$.
    \item \textbf{CL ALL}: The detector $\mathcal{M}$ is first trained on $\mathbb{D}_0$ and then retrained on each subsequent dataset using a continual learning technique that updates all its weights.
    \item \textbf{CL $\mathcal{M}_\text{E}$}: The detector $\mathcal{M}$ is initially trained on $\mathbb{D}_0$ and then retrained on the following datasets using a continual learning technique that only updates the weights of $\mathcal{M}_\text{E}$ while keeping those of $\mathcal{M}_\text{C}$ frozen.
    \item \textbf{CL $\mathcal{M}_\text{C}$}: It approach is the reverse of CL $\mathcal{M}_\text{E}$. The detector $\mathcal{M}$ is retrained on the datasets $\mathbb{D}_1$, $\mathbb{D}_2$, $\mathbb{D}_3$ using a continual learning technique that only updates the weights of $\mathcal{M}_\text{C}$ while keeping those of $\mathcal{M}_\text{E}$ frozen.
\end{itemize}

The goal of our analysis is to understand which of these training strategies yields the best results across different datasets.
Train-on-All represents the ideal case, as it has simultaneous access to all the considered datasets and offers the best possible outcome, while Fine-tuning serves as the lower bound due to its vulnerability to catastrophic forgetting~\cite{zhang2023you}.
As for the continual learning-based strategies, retraining all the model weights represents the standard approach considered in the literature, while updating only one module at a time is the novel scenario we aim to investigate.
The rationale behind this experiment is the following.
If we assume, as mentioned above, that the initial layers of the network are responsible for creating a meaningful low-dimensional representation of the input, we can imagine that updating only the $\mathcal{M}_\text{E}$ module may suffice to provide meaningful representations across different datasets, which can be effectively classified by the frozen $\mathcal{M}_\text{C}$ module.
On the other hand, if the $\mathcal{M}_\text{E}$ encoder is already capable of extracting a meaningful representation from the input data, it can be kept frozen, while retraining the $\mathcal{M}_\text{C}$ classifier to learn how to discriminate between features coming from different datasets.
In general, by limiting the number of weights updated during retraining, we reduce the risk of the model forgetting previously learned knowledge, which is the primary goal of continual learning.

The continual learning strategy we consider is \gls{dfwf}, proposed in~\cite{ma2021continual}.
This technique has been proposed explicitly for fake audio detection and addresses the challenge of adapting the model to new spoofing attacks while retaining knowledge of previously learned ones. 
\gls{dfwf} incorporates two key constraints: \gls{LwF}, a knowledge distillation loss that preserves the model's prior knowledge, and \gls{PSA}, a distance measure that maintains consistency in recognizing authentic audio.
By combining these two strategies, \gls{dfwf} mitigates catastrophic forgetting, allowing the model to learn new attack types without sacrificing its ability to identify previously encountered ones.
We refer the reader to the original paper~\cite{ma2021continual} for additional details.


\section{Experimental Setup}
\label{sec:setup}

In this section, we present the experimental setup we utilized in our analyses.
We start by introducing the datasets that we considered during both the training and test phases of the models.
Then, we provide details about the detectors we used, detailing the parameters and the \gls{dfwf} loss used to train them.

\subsection{Datasets}

To effectively evaluate the knowledge retention capabilities of the models trained using the proposed continual learning methods, we considered four different speech deepfake detection datasets.
We employed these datasets in the same order as they were presented in the literature.
We did so to mimic a real-world scenario, where a speech deepfake detector trained on an older dataset must be retrained to address new spoofing techniques, without losing the knowledge acquired from the previous data.
All audio data were sampled at a frequency of $f_\text{s} = \SI{16}{\kilo\hertz}$.

\noindent \textbf{ASVspoof 2019~\cite{todisco2019asvspoof}.}
Released for the homonymous challenge, this dataset was designed to develop effective ASV models.
It includes real speech from the VCTK corpus~\cite{veaux2016superseded} and synthetic speech generated by \num{19} different algorithms, with data distributed unevenly across the train, development, and evaluation sets to enable open-set evaluation.

\noindent \textbf{FakeOrReal~\cite{reimao2019dataset}.}
This dataset contains over \num{198000} utterances of both real and synthetic speech. The synthetic data were generated using various \gls{tts} systems, including both open-source and commercial tools. The real data are sourced from a diverse range of open-source speech datasets and additional sources such as TED Talks and YouTube videos.

\noindent \textbf{In-the-Wild~\cite{muller2022does}.} 
This dataset aims to evaluate speech deepfake detectors in realistic environments by providing in-the-wild speech data. It consists of nearly \num{38} hours of audio, evenly split between fake and real speech. The fake clips were created by segmenting publicly accessible video and audio files, while the real clips come from publicly available material featuring the same speakers.

\noindent \textbf{Purdue speech dataset~\cite{bhagtani2024recent}.} 
This corpus includes \num{25000} synthetic speech tracks generated by five advanced diffusion model-based voice cloning methods: ProDiff, DiffGAN-TTS, ElevenLabs, UnitSpeech, and XTTS. The real speech data were sourced from LJSpeech~\cite{LJSpeech} and LibriSpeech~\cite{panayotov2015librispeech}.

\subsection{Speech deepfake detectors}

In this study, we evaluate the proposed training approach considering two distinct speech deepfake detectors. We do so to verify whether the obtained results are specific to a single model or can be generalized.

\noindent \textbf{RawNet2~\cite{tak2021end}.}
This is an end-to-end neural network that operates on raw waveform inputs.
Its architecture includes Sinc filters~\cite{ravanelli2018speaker}, followed by two Residual Blocks with skip connections on top of a GRU layer to extract frame-level representations of the input signal.
For our experiments, the $\mathcal{M}_\text{E}$ module includes all the layers up to the GRU, while the $\mathcal{M}_\text{C}$ module comprises the subsequent ones.

\noindent \textbf{LCNN~\cite{wu2018light}.} 
This is a Light CNN model adapted to the speech deepfake detection task.
It operates on Mel spectrograms extracted from the input audio and integrates several convolutional layers with varying kernel sizes and strides to capture different levels of abstraction, together with multi-level feature mapping techniques to enhance feature extraction.
In our experiments, we define $\mathcal{M}_\text{E}$ to encompass the initial \num{5} convolutional layers of the model, while $\mathcal{M}_\text{C}$ comprises the final \num{4} convolutional layers and the subsequent dense layer.

\subsection{Training setup}

We trained all the models for \num{150} epochs by monitoring the validation loss value computed using the Cross-Entropy function.
We assumed a \num{10} epochs patience for the early stopping, a batch size of \num{128}, and an AdamW optimizer. The learning rate, initially set to \num{e-4}, was reduced according to a cosine annealing schedule.
During training, each batch was balanced to include an equal number of samples from the real and fake classes.

To compute the \gls{dfwf} loss, we considered the CrossEntropy loss, \gls{LwF}, and \gls{PSA} terms as equally important. This approach ensured a balanced model optimization by equally prioritizing classification accuracy, knowledge retention, and embedding alignment.

\section{Results}
\label{sec:results}

In this section, we evaluate the performance of the proposed continual learning strategies and investigate whether these can enhance knowledge retention in state-of-the-art speech deepfake detectors.

\begin{figure}
    \centering
    \includegraphics[width=0.7\columnwidth]{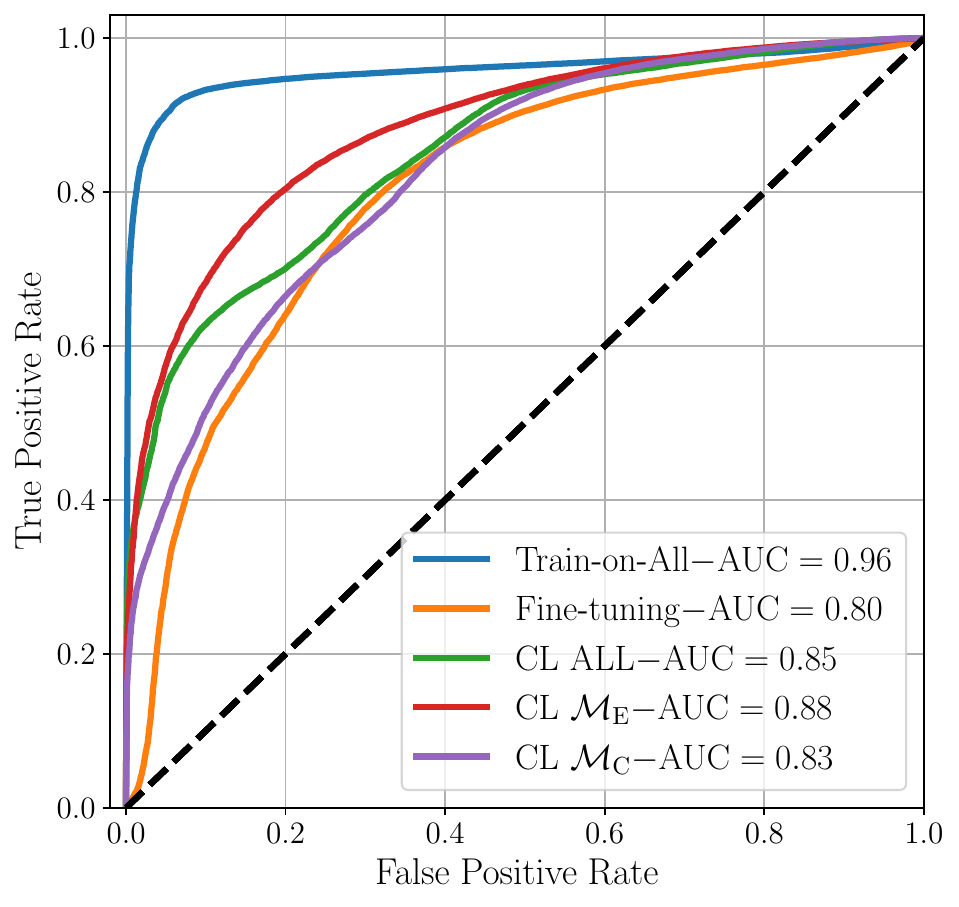}
    \caption{\gls{roc} curves of RawNet2 trained with different approaches and tested on all the considered datasets.}
    \label{fig:RawNet_ROC}
    \vspace{-1em}
\end{figure}

As a first experiment, we trained the RawNet2 and LCNN models using the five training strategies introduced in Section~\ref{sec:method} across the considered datasets and tested them on the union of all the sets.
The evaluation partition of each corpus is disjointed from the training and validation ones.
Figure~\ref{fig:RawNet_ROC} shows the results of this analysis for RawNet2 by means of \gls{roc} curves and AUC values.
Due to space constraints, the corresponding plot for the LCNN-based detector is not included as it shows the same behavior.
As anticipated, the best-performing approach is Train-on-All, which achieves an overall AUC value of 96\% with RawNet2 and 95\% with LCNN.
The reason behind this result is that this training approach provides the detector with simultaneous access to all the datasets, enabling it to learn their distributions effectively and enhance its training process.
Conversely, the worst-performing strategy for both models is fine-tuning, scoring an AUC of 80\% with RawNet2 and 81\% with LCNN.
Regarding the continual learning strategies, all of them outperform the Fine-tuning approach, highlighting the effectiveness of the \gls{dfwf} technique.
However, when considering the overall AUC value across all datasets, there is no significant difference in the performance of the three approaches.

To further investigate the performance differences among the continual learning approaches, we evaluated each model individually on every dataset.
Table~\ref{tab:results_rawnet} and Table~\ref{tab:results_LCNN} show the results of this analysis using balanced accuracy as the evaluation metric.
For a fair assessment across all datasets, we computed the balanced accuracy value by considering the optimal threshold based on the \gls{roc} curves calculated above.
As shown above, the best-performing approach is Train-on-All, while Fine-tuning performs the worst.
From the results shown here, the effect of catastrophic forgetting is evident. Although achieving 99\% accuracy with both detectors on the Purdue dataset, their accuracy drops significantly on datasets encountered in earlier training steps, with RawNet2 and LCNN scoring only 50\% and 52\% on the FakeOrReal corpus, respectively.

\begin{table}
\centering
\caption{Balanced accuracy values of the RawNet2 model across diverse datasets using different training strategies. 
}
\label{tab:results_rawnet}
\begin{tabular}{ccccc|c}
\toprule
  & ASV19 & FOR & ITW & Purdue & Avg. \\ \midrule \midrule
Train-on-All            & 89\%             & 85\%          & 98\%         & 99\%    & 93\%     \\ 
Fine-tuning               & 63\%             & 50\%          & 64\%         & 99\%    & 69\%     \\ \midrule
CL ALL                & 62\%             & 72\%          & 72\%         & \textbf{99\%}    & 76\%     \\
CL $\mathcal{M}_\text{E}$          & \textbf{85\%}             & \textbf{82\%}          & \textbf{81\%}         & 76\%    & \textbf{81\%}     \\ 
CL $\mathcal{M}_\text{C}$     & 62\%             & 68\%          & 69\%         & 96\%    & 74\%     \\ \bottomrule
\end{tabular}
\end{table}

\begin{table}
\centering
\caption{Balanced accuracy values of the MelSpec+LCNN model across diverse datasets using different training strategies. 
}
\label{tab:results_LCNN}
\begin{tabular}{ccccc|c}
\hline
\toprule
  & ASV19 & FOR & ITW & Purdue & Avg. \\ \midrule \midrule
Train-on-All            & 90\%             & 93\%          & 98\%         & 98\%    & 95\%     \\
Fine-tuning               & 67\%             & 52\%          & 66\%         & 99\%    & 71\%     \\ \midrule
CL ALL                & 78\%             & 57\%          & 86\%         & 97\%    & 80\%     \\ 
CL $\mathcal{M}_\text{E}$          & \textbf{81\%}             & 62\%          & \textbf{87\%}         & 94\%    & \textbf{81\%}     \\ 
CL $\mathcal{M}_\text{C}$     & 73\%             & \textbf{64\%}          & 74\%         & \textbf{98\%}    & 77\%     \\ \bottomrule
\end{tabular}
\vspace{-.5em}
\end{table}

This analysis also reveals significant performance differences between the three continual learning strategies that were not apparent in the previous experiment.
The \textit{CL ALL} strategy achieves acceptable accuracy values on the FakeOrReal, In-the-Wild, and Purdue datasets but performs poorly on ASVspoof2019, particularly with the RawNet2 model (bal. acc. 62\%).
This suggests signs of forgetting when multiple retraining steps are involved. A similar, but more pronounced behavior is observed with the \textit{CL} $\mathcal{M}_\text{C}$ strategy, which shows poor performance on the early training datasets for both RawNet2 and LCNN, making it the least effective approach among the continual learning methods we evaluated.
In contrast, \textit{CL} $\mathcal{M}_\text{E}$ demonstrates excellent knowledge retention capabilities, emerging as the best-performing among the continual learning strategies, with an average balanced accuracy value of 81\% for both RawNet2 and LCNN. 
Despite its excellent performance on the previously seen datasets, \textit{CL} $\mathcal{M}_\text{E}$ yields the lowest results on the Purdue dataset among the five training strategies we considered, with a balanced accuracy value of 76\% with RawNet2 and 94\% with LCNN.
We hypothesize that this may be due to the relatively low weight assigned to the CrossEntropy loss for the new dataset during training, compared to the \gls{LwF} and \gls{PSA} terms.
This likely enabled the model to retain significant knowledge from previous sets but limited its ability to learn effectively from the current dataset.

\begin{figure}
    \centering
    \includegraphics[width=0.75\columnwidth]{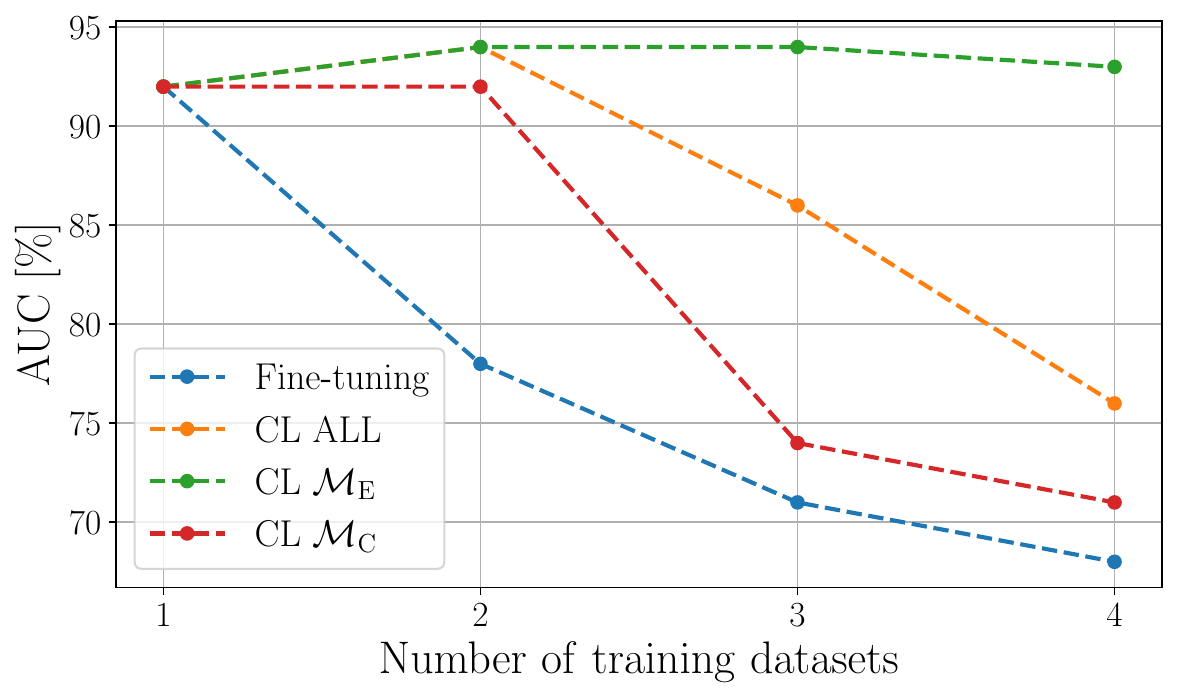}
    \caption{AUC value of the RawNet2 model on the ASVspoof 2019 dataset trained with different strategies, as a function of the number of datasets used for training.}
    \label{fig:RawNet_AUC_asvspoof}
\end{figure}

\begin{figure}
    \centering
    \includegraphics[width=0.75\columnwidth]{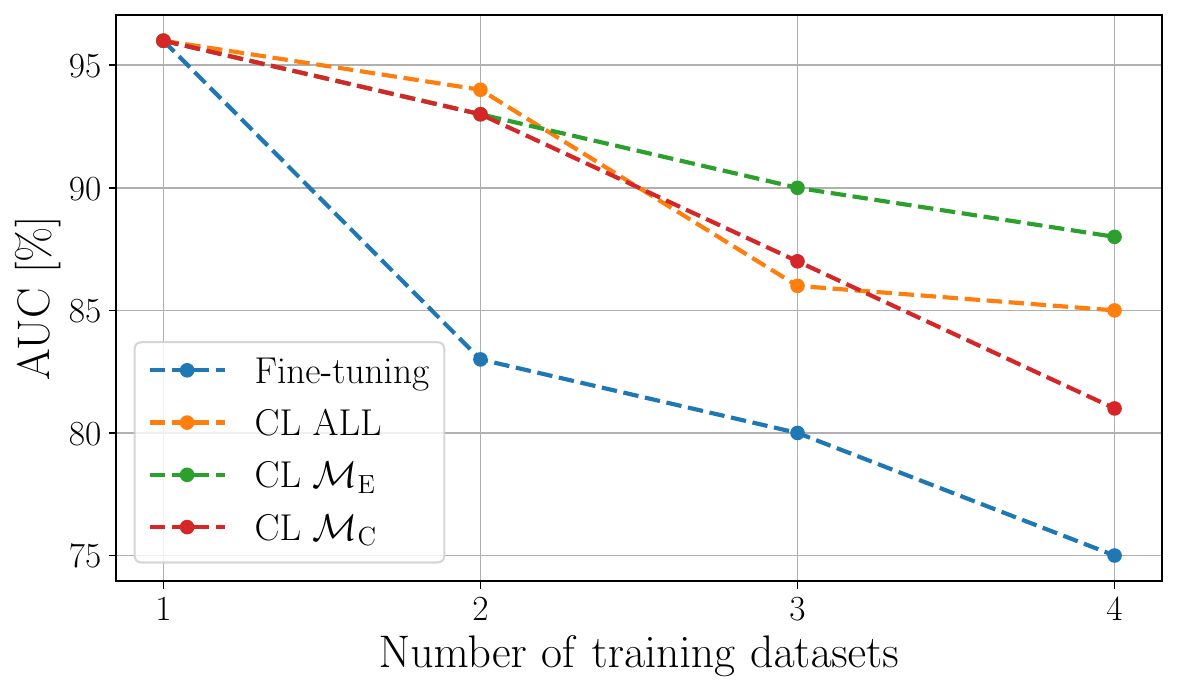}
    \caption{AUC value of the LCNN model on the ASVspoof 2019 dataset trained with different strategies, as a function of the number of datasets used for training.}
    \label{fig:LCNN_AUC_asvspoof}
    \vspace{-.5em}
\end{figure}

As a final experiment, we investigated how the knowledge of previous datasets diminishes as the number of training steps increases.
This study aims to evaluate how much each training strategy is prone to catastrophic forgetting.
Figure~\ref{fig:RawNet_AUC_asvspoof} and Figure~\ref{fig:LCNN_AUC_asvspoof} present the results of this analysis for the RawNet2 and LCNN models, respectively.
Results are shown in terms of AUC on ASVspoof 2019 when the training is performed using up to four datasets. 
When the number of datasets is equal to \num{1}, we only consider the ASVspoof 2019 dataset for training. As the number increases, we add an additional dataset to the training process, up to a total of \num{4} datasets (FakeOrReal, In-the-Wild, and Purdue).
The two figures reveal a similar pattern among the training strategies, with Fine-tuning being the lowest performer due to its high susceptibility to catastrophic forgetting and \textit{CL} $\mathcal{M}_\text{E}$ being the highest, demonstrating superior knowledge retention compared to other methods. Specifically, for RawNet2, its AUC value remains nearly constant across the training steps.
\section{Conclusions}
\label{sec:conclusion}

In this paper, we investigated the effectiveness of different continual learning approaches for training speech deepfake detectors, aiming to maintain knowledge while adapting models to new synthetic speech generation techniques.
Our experiments show that both the RawNet2 and LCNN models benefit from the proposed \textit{CL} $\mathcal{M}_\text{E}$ approach, suggesting that the most critical component of a speech deepfake detector is the part that analyzes the input features, rather than its final classifier.

Future work will focus on studying new continual learning techniques and strategies to mitigate catastrophic forgetting during model training, aiming to further improve knowledge retention and adaptability of the detectors to real-world scenarios.

\newpage

\bibliographystyle{IEEEtran}
\bibliography{bstcontrol.bib, biblio.bib}

\begin{thebibliography}{10}
\providecommand{\url}[1]{#1}
\csname url@samestyle\endcsname
\providecommand{\newblock}{\relax}
\providecommand{\bibinfo}[2]{#2}
\providecommand{\BIBentrySTDinterwordspacing}{\spaceskip=0pt\relax}
\providecommand{\BIBentryALTinterwordstretchfactor}{4}
\providecommand{\BIBentryALTinterwordspacing}{\spaceskip=\fontdimen2\font plus
\BIBentryALTinterwordstretchfactor\fontdimen3\font minus \fontdimen4\font\relax}
\providecommand{\BIBforeignlanguage}[2]{{%
\expandafter\ifx\csname l@#1\endcsname\relax
\typeout{** WARNING: IEEEtran.bst: No hyphenation pattern has been}%
\typeout{** loaded for the language `#1'. Using the pattern for}%
\typeout{** the default language instead.}%
\else
\language=\csname l@#1\endcsname
\fi
#2}}
\providecommand{\BIBdecl}{\relax}
\BIBdecl

\bibitem{cuccovillo2022open}
L.~Cuccovillo, C.~Papastergiopoulos, A.~Vafeiadis, A.~Yaroshchuk, P.~Aichroth, K.~Votis, and D.~Tzovaras, ``Open challenges in synthetic speech detection,'' in \emph{IEEE International Workshop on Information Forensics and Security (WIFS)}, 2022.

\bibitem{amerini2024deepfake}
I.~Amerini, M.~Barni, S.~Battiato, P.~Bestagini, G.~Boato, T.~S. Bonaventura, V.~Bruni, R.~Caldelli, F.~De~Natale, R.~De~Nicola \emph{et~al.}, ``{Deepfake Media Forensics: State of the Art and Challenges Ahead},'' in \emph{International Workshop on Safeguarding Social Networks (SAFE-SN)}, 2024.

\bibitem{yang2024robust}
Y.~Yang, H.~Qin, H.~Zhou, C.~Wang, T.~Guo, K.~Han, and Y.~Wang, ``A robust audio deepfake detection system via multi-view feature,'' in \emph{IEEE International Conference on Acoustics, Speech and Signal Processing (ICASSP)}, 2024.

\bibitem{xie2023domain}
Y.~Xie, H.~Cheng, Y.~Wang, and L.~Ye, ``Domain generalization via aggregation and separation for audio deepfake detection,'' \emph{IEEE Transactions on Information Forensics and Security}, 2023.

\bibitem{li2023voice}
L.~Li, T.~Lu, X.~Ma, M.~Yuan, and D.~Wan, ``Voice deepfake detection using the self-supervised pre-training model hubert,'' \emph{Applied Sciences}, vol.~13, no.~14, p. 8488, 2023.

\bibitem{mari2023all}
D.~Mari, D.~Salvi, P.~Bestagini, and S.~Milani, ``{All-for-One and One-For-All: Deep learning-based feature fusion for Synthetic Speech Detection},'' in \emph{European Conference on Machine Learning and Knowledge Discovery in Databases Workshop (ECML/PKDD)}, 2023.

\bibitem{bhagtani2024recent}
K.~Bhagtani, A.~K.~S. Yadav, P.~Bestagini, and E.~J. Delp, ``{Are Recent Deepfake Speech Generators Detectable?}'' in \emph{ACM Workshop on Information Hiding and Multimedia Security}, 2024.

\bibitem{patel2023deepfake}
Y.~Patel, S.~Tanwar, R.~Gupta, P.~Bhattacharya, I.~E. Davidson, R.~Nyameko, S.~Aluvala, and V.~Vimal, ``{Deepfake generation and detection: Case study and challenges},'' \emph{IEEE Access}, 2023.

\bibitem{ma2021continual}
H.~Ma, J.~Yi, J.~Tao, Y.~Bai, Z.~Tian, and C.~Wang, ``Continual learning for fake audio detection,'' in \emph{Conference of the International Speech Communication Association (INTERSPEECH)}, 2021.

\bibitem{zhang2023you}
X.~Zhang, J.~Yi, J.~Tao, C.~Wang, and C.~Y. Zhang, ``{Do you remember? Overcoming catastrophic forgetting for fake audio detection},'' in \emph{International Conference on Machine Learning (ICML)}, 2023.

\bibitem{zhang2024remember}
X.~Zhang, J.~Yi, C.~Wang, C.~Y. Zhang, S.~Zeng, and J.~Tao, ``{What to remember: Self-adaptive continual learning for audio deepfake detection},'' in \emph{AAAI Conference on Artificial Intelligence}, 2024.

\bibitem{wang2022low}
C.~Wang, J.~Yi, X.~Zhang, J.~Tao, X.~Yan, L.~Xu, and R.~Fu, ``{Low-rank Adaptation Method for Wav2vec2-based Fake Audio Detection},'' in \emph{Workshop on Deepfake Audio Detection and Analysis (DADA)}, 2022.

\bibitem{mulimani2024class}
M.~Mulimani and A.~Mesaros, ``{Class-Incremental Learning for Multi-Label Audio Classification},'' in \emph{IEEE International Conference on Acoustics, Speech and Signal Processing (ICASSP)}, 2024.

\bibitem{dong2024advancing}
F.~Dong, Q.~Tang, Y.~Bai, and Z.~Wang, ``{Advancing Continual Learning for Robust Deepfake Audio Classification},'' \emph{arXiv preprint arXiv:2407.10108}, 2024.

\bibitem{todisco2019asvspoof}
M.~Todisco, X.~Wang, V.~Vestman, M.~Sahidullah, H.~Delgado, A.~Nautsch, J.~Yamagishi, N.~Evans, T.~Kinnunen, and K.~A. Lee, ``{ASVspoof 2019: Future horizons in spoofed and fake audio detection},'' in \emph{Conference of the International Speech Communication Association (INTERSPEECH)}, 2019.

\bibitem{veaux2016superseded}
C.~Veaux, J.~Yamagishi, K.~MacDonald \emph{et~al.}, ``{Superseded-CSTR VCTK Corpus: English Multi-Speaker Corpus for CSTR Voice Cloning Toolkit},'' \emph{University of Edinburgh. The Centre for Speech Technology Research (CSTR)}, 2016.

\bibitem{reimao2019dataset}
R.~Reimao and V.~Tzerpos, ``{FOR: A dataset for synthetic speech detection},'' in \emph{IEEE International Conference on Speech Technology and Human-Computer Dialogue (SpeD)}, 2019.

\bibitem{muller2022does}
N.~M. M{\"u}ller, P.~Czempin, F.~Dieckmann, A.~Froghyar, and K.~B{\"o}ttinger, ``Does audio deepfake detection generalize?'' in \emph{Conference of the International Speech Communication Association (INTERSPEECH)}, 2022.

\bibitem{LJSpeech}
K.~Ito and L.~Johnson, ``{The LJ Speech Dataset},'' \url{https://keithito.com/LJ-Speech-Dataset/}, 2017.

\bibitem{panayotov2015librispeech}
V.~Panayotov, G.~Chen, D.~Povey, and S.~Khudanpur, ``Librispeech: an {ASR} corpus based on public domain audio books,'' in \emph{IEEE International Conference on Acoustics, Speech and Signal Processing (ICASSP)}, 2015.

\bibitem{tak2021end}
H.~Tak, J.~Patino, M.~Todisco, A.~Nautsch, N.~Evans, and A.~Larcher, ``{End-to-end anti-spoofing with RawNet2},'' in \emph{IEEE International Conference on Acoustics, Speech and Signal Processing (ICASSP)}, 2021.

\bibitem{ravanelli2018speaker}
M.~Ravanelli and Y.~Bengio, ``{Speaker Recognition from Raw Waveform with SincNet},'' in \emph{IEEE Spoken Language Technology Workshop (SLT)}, 2018.

\bibitem{wu2018light}
X.~Wu, R.~He, Z.~Sun, and T.~Tan, ``{A light CNN for deep face representation with noisy labels},'' \emph{{IEEE Transactions on Information Forensics and Security}}, vol.~13, no.~11, pp. 2884--2896, 2018.

\end{thebibliography}

\end{document}